\documentclass[letterpaper,10pt]{article}

\usepackage{graphicx}
\usepackage{amsmath}
\usepackage{amssymb}
\usepackage{psfrag}
\newcommand{\dirac}{{\slash \negthinspace \negthinspace \negthinspace \nabla}}
\newcommand{\dd}{\textrm{d}}

\title{Quasinormal frequencies of asymptotically flat two-dimensional black holes}

\author{A.\ L\'opez-Ortega\thanks{alopezo ``at'' ipn.mx} $^1$, I.\ Vega-Acevedo\thanks{ivega ``at'' esfm.ipn.mx} $^2$  \\
		$^1$ Centro de Investigaci\'on en Ciencia Aplicada y Tecnolog\'{\i}a Avanzada. \\
	      Unidad Legaria. Instituto Polit\'ecnico Nacional. \\
             Calzada Legaria \# 694. Colonia Irrigaci\'on. Delegaci\'on Miguel Hidalgo. \\
	      M\'exico, D.\ F., M\'exico. \\
	      C.\ P.\  11500\\   
       $^2$ Departamento de F\'{\i}sica. Escuela Superior de F\'{\i}sica y Matem\'aticas. \\
Instituto Polit\'ecnico Nacional. \\
Unidad Profesional Adolfo L\'opez Mateos, Edificio 9. \\
M\'exico, D.\ F., M\'exico. \\
C.\ P.\ 07738 }

\begin{document}

\maketitle

\begin{abstract}

We discuss whether the minimally coupled massless Klein-Gordon and Dirac fields have well defined quasinormal modes in  single horizon, asymptotically flat two-di\-men\-sion\-al black holes. To get the result we solve the equations of motion in the massless limit and we also calculate the effective potentials of Schr\"odinger type equations. Furthermore we calculate exactly the quasinormal frequencies of the Dirac field propagating in the two-dimensional uncharged Witten black hole. We compare our results on its quasinormal frequencies with other already published.



\end{abstract}

\section{Introduction}
\label{s: Introduction}

Two-dimensional gravity theories are widely studied \cite{Grumiller:2002nm}. Although the simplifications in two-dimensional gravitational models remove some relevant features of higher dimensional gravitational systems, these also allow more detailed analysis of some physical phenomena. Furthermore these simplifications allow to address some conceptual problems in a simple framework \cite{Grumiller:2002nm}, \cite{Grumiller:2006rc}. Thus among the reasons for analyzing two-dimensional gravity models we enumerate:

\begin{description}
\item[i)]  The models of two-dimensional gravity appear as effective gravity theories in string theory, in spherically symmetric or axial reduction of higher dimensional gravitational systems, and as a limit case in other physical systems  \cite{Grumiller:2002nm}.

\item[ii)] These two-dimensional models reproduce qualitatively many features of the physical phenomena of higher dimensional systems as Hawking evaporation, black hole thermodynamics, and critical collapse \cite{Grumiller:2002nm}, \cite{Grumiller:2006rc}.

\item[iii)]  Two-dimensional gravity is an appropriate framework for testing some ideas about quantum gravity \cite{Grumiller:2002nm}.
\end{description}

Two-dimensional gravity theories have black hole solutions \cite{Grumiller:2002nm} and these two-di\-men\-sion\-al solutions are useful to test some ideas on how to solve  several problems that we find in the physics of black holes, as the origin of their entropies and the information loss problem \cite{Grumiller:2002nm}. Taking into account the usefulness of these two-dimensional black hole solutions, it is appropriate to study other aspects of their behavior, for example, how these two-dimensional black holes react when we slightly perturb them with a test field. 

It is well known that when we perturb a black hole with a test field, it oscillates with well defined modes that satisfy the appropriate radiation boundary conditions at the horizon and at the asymptotic region. These decaying modes are called quasinormal modes (QNM in what follows) and their complex frequencies are named quasinormal frequencies (QNF in what follows) \cite{Kokkotas:1999bd}, \cite{Berti:2009kk}.

Near the horizon the QNM boundary condition usually imposed is that the field must be purely ingoing. This condition arises from the fact that classical black holes do not emit radiation. Far from the event horizon the QNM boundary condition that the field must satisfy depends on the asymptotic structure of the black hole. For example, for asymptotically flat black holes we impose that the field must be purely outgoing far from the event horizon, whereas for asymptotically anti-de Sitter black holes we usually impose that the field (or its flux) goes to zero at infinity \cite{Kokkotas:1999bd}, \cite{Berti:2009kk}.

The computations of the QNF are useful to know some properties of the black hole as for example its classical stability \cite{Kokkotas:1999bd}, \cite{Berti:2009kk} and the determination of its entropy quanta \cite{Hod:1998vk}--\cite{Maggiore:2007nq}. Furthermore since the QNF depend on the physical parameters of the black hole as its mass, electric charge, and angular momentum, they give us a tool for determining these physical parameters. Considering the relevance of the two-dimensional black holes \cite{Grumiller:2002nm}, \cite{Grumiller:2006rc}, we think that an useful addition to the known results is the calculation of their QNF.

Taking into account the simplicity of lower dimensional systems we expect that for some lower dimensional black holes their QNF can be calculated exactly. Recently appeared several examples of lower dimensional black holes whose QNF are exactly computed, for example, we are aware of the gravitational systems studied in \cite{Birmingham:2001pj}--\cite{Becar:2010zz}. In these references are calculated exactly the QNF for the three-dimensional BTZ black hole \cite{Birmingham:2001pj}--\cite{Crisostomo:2004hj}, for the three-dimensional dilatonic black holes \cite{Fernando:2003ai}-\cite{Fernando:2009tv}, for the three-dimensional warped AdS(3) black holes \cite{Chen:2009hg}-\cite{Li:2010sv}, for the three-dimensional de Sitter spacetime \cite{Du:2004jt}, \cite{Lopez-Ortega:2006ig}, for the two-dimensional de Sitter spacetime \cite{Zelnikov:2008rg}, and for the two-dimensional Witten black hole \cite{Zelnikov:2008rg}--\cite{Becar:2010zz}. 

Furthermore in two dimensions we expect to find relevant results  for a family of black hole solutions. For example,  in \cite{Zelnikov:2008rg} are calculated exactly the QNF for a family of two-dimensional black holes. We also note that in \cite{Kettner:2004aw} Kettner et al.\ compute the asymptotic QNF of the Klein-Gordon field non-minimally coupled to the dilaton propagating in a family of black holes that are solutions to the equations of motion obtained from actions with power law potentials \cite{Cadoni:1995dd}. 

Here we investigate whether the massless Klein-Gordon and Dirac fields have well defined QNF in single horizon, asymptotically flat two-dimensional black holes. Furthermore we calculate exactly the QNF of a test Dirac field in Witten black hole. The results about the QNF of the Dirac field in Witten black hole extend those obtained for the Klein-Gordon field \cite{Becar:2007hu}, \cite{LopezOrtega:2009zx} and the massless Dirac field \cite{Becar:2010zz}.

This paper is organized as follows. In Section \ref{s: QNM massless} we analyze whether the massless Klein-Gordon and Dirac equations have solutions that satisfy the boundary conditions of the QNM for single horizon, asymptotically flat two dimensional black holes. For both fields we also calculate the effective potentials of Schr\"odinger type equations and use these potentials to discuss the existence of their QNF. In Section \ref{s: Witten black hole} we exactly calculate the QNF of the Dirac field propagating in the two-dimensional Witten black hole. We also compare our results with other already published on the QNF of this two-dimensional black hole. Finally in Section \ref{s: Summary} we summarize and discuss the obtained results.

\section{Quasinormal modes of massless fields in two-dimensional black holes}
\label{s: QNM massless}

For a generic single horizon, asymptotically flat two-dimensional black hole in this Section we study whether the massless Klein-Gordon and Dirac equations have solutions that satisfy the boundary conditions of the QNM. In what follows we consider two-dimensional black holes with line element
\begin{equation} \label{eq: metric general}
\dd s^{2}=f\dd t^{2}-\frac{\dd r^{2}}{g},
\end{equation}
with $f$ and $g$ being functions of the $r$ coordinate. We assume that the functions $f$ and $g$ allow the existence of an event horizon at $r=r_H$ and when $r \to + \infty$ the spacetime is asymptotically flat. 

Furthermore we define the appropriate tortoise coordinate $x$ for spacetime (\ref{eq: metric general}) by
\begin{equation}
\label{eq: tortoise coordinate general}
 x = \int \frac{\dd r}{\sqrt{f g}},
\end{equation} 
and we assume that it satisfies
\begin{eqnarray} \label{eq: conditions tortoise}
 \lim_{r \to r_H} x \to - \infty , \qquad \quad  \lim_{r  \to + \infty} x \to + \infty,
\end{eqnarray} 
thus $x \in (-\infty, +\infty)$ as $r \in (r_H, +\infty)$. 

An example of black hole that satisfies the conditions that we have imposed on the background is the two-dimensional Schwarzschild black hole for which
\begin{equation}
 f(r) = g(r) = 1 - \frac{2 M}{r},
\end{equation} 
where $r$ is the radial coordinate, $M$ is related to the mass of the black hole, and it has an event horizon at $r_H = 2 M$. From formula (\ref{eq: tortoise coordinate general}) we find that for the two-dimensional Schwarzschild black hole the tortoise coordinate is equal to
\begin{equation}
 x = r + 2 M \ln(r -2 M),
\end{equation} 
and we note that it satisfies conditions (\ref{eq: conditions tortoise}).

For these black holes we define the QNM of test fields as the oscillations that satisfy the boundary conditions:
\begin{description}
\item[a)] The oscillations are purely ingoing near the event horizon.

\item[b)] The oscillations are purely outgoing at infinity.
\end{description}

Let us begin with the  minimally coupled Klein-Gordon field that for mass different from zero satisfies the equation
\begin{equation} \label{eq: Klein-Gordon}
\left(\square + m^{2}\right) \Phi = 0 ,
\end{equation}
where $\square$ denotes the two-dimensional d'Alembertian and $m$ is the mass of the Klein-Gordon field. For solving  the  Klein-Gordon equation in the two-dimensional metric (\ref{eq: metric general}) we propose a separable solution of the form 
\begin{equation}
\label{eq: separation KG}
\Phi(r,t) = R(r)e^{-i\omega t},
\end{equation}
to find that Eq.\ (\ref{eq: Klein-Gordon}) reduces to the radial differential equation
\begin{equation} \label{eq: KG massive radial}
  \frac{\omega^2}{f} R   + \sqrt{\frac{g}{f}}\frac{\dd }{\dd r} \left( \sqrt{gf} \frac{\dd R}{\dd r} \right)  - m^2 R = 0.
\end{equation} 

Using the tortoise coordinate (\ref{eq: tortoise coordinate general}) for the line element (\ref{eq: metric general}) we get that in the massless limit Eq.\ (\ref{eq: KG massive radial}) may be written as
\begin{equation} \label{eq: equation Schrodinger massless}
\frac{\dd^{2} R}{\dd x^{2}}+\omega^{2} R = 0,
\end{equation}
whose solutions are 
\begin{equation} \label{eq: radial massless KG}
 R = A e^{i \omega x} + B e^{- i \omega x},
\end{equation} 
where $A$ and $B$ are constants.

If the tortoise coordinate satisfies the previously enumerated conditions in formulas (\ref{eq: conditions tortoise}), then the first term in (\ref{eq: radial massless KG}) is an outgoing wave near the event horizon and at infinity. The second term in formula (\ref{eq: radial massless KG}) is an ingoing wave near the event horizon and at infinity. Thus the solutions (\ref{eq: radial massless KG}) can not satisfy the boundary conditions of the QNMs for asymptotically flat black holes and we obtain that the massless Klein-Gordon field does not have well defined QNM in a single horizon, asymptotically flat two-dimensional black hole.

Now we study the massless Dirac field. We first note that in metric (\ref{eq: metric general}) the Dirac equation reduces to a pair of coupled partial differential equations and that these equations decouple in the massless limit. To see this fact we write the Dirac equation
\begin{equation} \label{eq: Dirac equation}
i \dirac \Psi= m \Psi,
\end{equation}
in the two-dimensional metric (\ref{eq: metric general}) and use the method of Ref.\ \cite{LopezOrtega:2009qc} to reduce the Dirac equation to ordinary differential equations. As usual, $\Psi$ is a two-spinor, $\dirac$ denotes the Dirac operator, and $m$ is the mass of the fermion field. The Dirac operator is given by $\dirac = \gamma^\mu \nabla_\mu$ where $\nabla_\mu$ denotes the covariant derivative, the matrices $\gamma^\mu$ satisfy $\gamma^\mu \gamma^\nu + \gamma^\nu \gamma^\mu = 2 g^{\mu \nu}$, and in two dimensions they are square $ 2 \times 2$ matrices. See \cite{Becar:2010zz}, \cite{Mann:1991md}, \cite{Morsink:1991qr} for other studies on the Dirac field propagating in two-dimensional spacetimes.

To simplify the Dirac equation first we factor out the function $f$ in metric (\ref{eq: metric general}). Thus
\begin{equation} \label{eq: metric factor}
\dd s^{2} = f \left( \dd t^{2}-\frac{ \dd r^{2} }{ f g} \right) = f \dd \tilde{s}^2,
\end{equation}
where
\begin{equation} \label{eq: reduced metric}
 \dd \tilde{s}^2 = \dd t^{2}-\frac{ \dd r^{2} }{ f g} .
\end{equation} 
Note that under a conformal transformation of metric (\ref{eq: metric general}), that is
\begin{equation}
 g_{\mu \nu} = f \tilde{g}_{\mu \nu},
\end{equation} 
if we change the two spinor, the Dirac operator, and the mass in the form 
\begin{align} \label{eq: transformations}
\Psi=\frac{1}{f^{1/4}} \tilde{\Psi} ,     \qquad \quad
\dirac \Psi = \frac{1}{f^{3/4}} \tilde{\dirac} \tilde{\Psi},   \qquad \quad
m=\frac{1}{f^{1/2}} \tilde{m}, 
\end{align} 
then the quantities $\tilde{\Psi}$, $\tilde{\dirac} \tilde{\Psi}$, and $\tilde{m}$ satisfy the Dirac equation in spacetime (\ref{eq: reduced metric}) \cite{LopezOrtega:2009qc}. 

Using the tortoise coordinate (\ref{eq: tortoise coordinate general}) we get that metric (\ref{eq: reduced metric}) takes the form
\begin{equation}
\label{eq: metric Minkowski}
\dd \tilde{s}^{2} = \dd t^{2} - \dd x^{2},
\end{equation}
and exploiting the chiral representation for the $\gamma^\mu$ matrices 
\begin{align} \label{eq: gamma matrices}
 \gamma^{0} = \gamma^{t} = \left( \begin{array}{cc}
0 & 1 \\
1 & 0 \end{array} \right),  \qquad \quad
 \gamma^{1}= \gamma^{r} =\left( \begin{array}{cc} 
0 &-1 \\ 
1 & 0 \end{array} \right),
\end{align} 
we find that the Dirac equation reduces to the pair of coupled partial differential equations
 \begin{align} \label{eq: Dirac coupled}
\partial_{t} \tilde{\Psi}_{2} - \partial_{x} \tilde{\Psi}_{2}= -i m \sqrt{f} \tilde{\Psi}_{1}, \nonumber \\
\partial_{t} \tilde{\Psi}_{1}  + \partial_{x} \tilde{\Psi}_{1}= -i m \sqrt{f} \tilde{\Psi}_{2} ,
\end{align} 
for the components $\tilde{\Psi}_{1}$ and $\tilde{\Psi}_{2}$ of the two-dimensional spinor
\begin{equation} \label{eq: two spinor} 
 \tilde{\Psi} = \left(\begin{array}{c}
\tilde{\Psi}_{1}\\
\tilde{\Psi}_{2}  \end{array} \right).
\end{equation} 

In the massless limit the previous system of coupled equations simplifies to the decoupled equations
\begin{align} \label{eq: massless limit Dirac}
\partial_{t} \tilde{\Psi}_{2} - \partial_{x} \tilde{\Psi}_{2}= 0, \nonumber \\
\partial_{t} \tilde{\Psi}_{1}  + \partial_{x} \tilde{\Psi}_{1}= 0 ,
\end{align}
or written in an equivalent form
\begin{align} \label{eq: massless limit Dirac 2}
\partial_{t} \tilde{\Psi}_{2} -  \sqrt{f g} \partial_{r}  \tilde{\Psi}_{2}= 0, \nonumber \\
\partial_{t} \tilde{\Psi}_{1}  + \sqrt{f g} \partial_{r}  \tilde{\Psi}_{1}= 0 .
\end{align}

Taking the components $\tilde{\Psi}_{1}$ and $\tilde{\Psi}_{2}$ of the two spinor as
\begin{equation} \label{eq: Dirac ansatz}
\tilde{\Psi}_{s}(x,t) = R_{s}(x)e^{-i\omega t},  \qquad \quad s=1,2,
\end{equation}
we obtain that Eqs.\ (\ref{eq: massless limit Dirac}) transform into
\begin{align}
\frac{\dd R_{2}}{\dd x} + i\omega R_{2} &=0,\nonumber\\
\frac{\dd R_{1}}{\dd x} -i\omega R_{1} &=0,
\end{align}
whose solutions are
\begin{align} \label{eq: solutions tortoise}
R_{1} = A e^{i\omega x} ,  \qquad \qquad 
R_{2} = B e^{-i\omega x},
\end{align}
with $A$ and $B$ constants.

Thus in the massless limit the function $R_1$ represents a purely outgoing wave near the horizon and far from the horizon, in a similar way the function $R_2$ represents a purely ingoing wave near the horizon and far from the horizon. Therefore the massless Dirac field can not satisfy the boundary conditions for the QNM of generic single horizon, asymptotically flat two-dimensional black holes.

To get these results we only assume that the two-dimensional black holes (\ref{eq: metric general}) have a single horizon, they are asymptotically flat, and their tortoise coordinates (\ref{eq: tortoise coordinate general}) satisfy conditions (\ref{eq: conditions tortoise}).

The previous results show that in single horizon, asymptotically flat two dimensional black holes, the massless Klein-Gordon and Dirac fields do not have modes that satisfy the boundary conditions of the QNM. Notice that the previous result is valid only for the QNM that satisfy the conditions a) and b) and that for the two dimensional black holes whose QNM satisfy other boundary conditions, we may find that the massless Klein-Gordon and Dirac fields have well defined QNF frequencies. Furthermore this result points out that  in two dimensional black holes it is necessary to analyze in detail the process of calculation for the QNF of the massive and massless fields, because it is possible that the QNF of the massless field can not be obtained  from the QNF of the massive field by a limiting procedure. 

We notice that an equivalent result is obtained for the Dirac and Klein-Gordon fields propagating in the three-dimensional extreme BTZ black hole \cite{Crisostomo:2004hj}. Moreover a similar result is true for the Klein-Gordon field that propagates in a five-dimensional dilatonic black hole. For this spacetime in Ref.\ \cite{LopezOrtega:2009zx} it is shown that for the $s$ wave of the massless Klein-Gordon field we can not satisfy the boundary conditions of the QNM. 

For single horizon, asymptotically flat two-dimensional black holes that are solutions to the  equations of motion derived from actions with power law potentials \cite{Cadoni:1995dd}, in Ref.\ \cite{Kettner:2004aw} Kettner et al.\ calculate the asymptotic QNF of the massless Klein-Gordon field. They find well defined asymptotic QNF and their conclusions do not contradict the previous results because in \cite{Kettner:2004aw} they calculate the asymptotic QNF of the massless Klein-Gordon field non-minimally coupled to the dilaton field and here we analyze the minimally coupled Klein-Gordon field. 

To support our previous results, in what follows we present the effective potentials \cite{Chandrasekhar book} for the radial equations of the Klein-Gordon and Dirac fields propagating in two dimensional spacetime with metric (\ref{eq: metric general}). For the massive Klein-Gordon field we note that the radial differential equation (\ref{eq: KG massive radial}) may be transformed into
\begin{equation} \label{eq: KG radial tortoise massive}
\frac{\dd^{2} R }{\dd x^{2}} + \omega^{2}  R = V R,
\end{equation}
where the effective potential $V$ is equal to
\begin{equation} \label{eq: potential KG}
V= m^{2} f.
\end{equation}

For the Dirac field, if we make the ansatz (\ref{eq: Dirac ansatz}), then we find that the system of coupled differential equations (\ref{eq: Dirac coupled}) reduces to
\begin{align} \label{eq: Dirac coupled massive}
&\frac{\dd R_{2} }{\dd x} + i\omega R_{2} = i\sqrt{f}mR_{1},\nonumber\\
&\frac{\dd R_{1} }{\dd x} - i\omega R_{1} =-i\sqrt{f}mR_{2}.
\end{align}
Changing the functions $R_1$ and $R_2$ by
\begin{align}
 R_1  = e^{i \pi /4} \tilde{R}_1,  \qquad \qquad R_2   = e^{- i \pi /4} \tilde{R}_2, 
\end{align} 
we obtain that the new functions $\tilde{R}_1$ and $\tilde{R}_2$ satisfy the coupled differential equations
\begin{align} \label{eq: Dirac coupled potential}
&\frac{\dd \tilde{R}_{2} }{\dd x} + i\omega \tilde{R}_{2} = - \sqrt{f}m \tilde{R}_{1},\nonumber\\
&\frac{\dd \tilde{R}_{1} }{\dd x} - i\omega \tilde{R}_{1} = - \sqrt{f} m \tilde{R}_{2} .
\end{align}

Following Chandrasekhar \cite{Chandrasekhar book} we define 
\begin{align}
 W  = - m \sqrt{f}, \qquad \qquad 
Z_{\pm}  = \tilde{R}_{1} \pm \tilde{R}_{2} ,
\end{align}
to find that the functions $Z_{\pm}$ satisfy Schr\"odinger type equations
\begin{equation} \label{eq: Dirac Schrodinger}
\frac{\dd^{2}Z_{\pm} }{\dd x^{2}} + \omega^{2} Z_{\pm} = V_{\pm} Z_{\pm} ,
\end{equation}
where
\begin{equation} \label{eq: potentials Dirac}
V_{\pm} = W^{2} \pm \frac{\dd W}{\dd x} = m^{2}f \mp \frac{m}{2} \sqrt{g}\frac{\dd f }{\dd r} = V \mp \frac{m}{2} \sqrt{g}\frac{\dd f }{\dd r}.
\end{equation}

Notice that for the massive Klein-Gordon field the effective potential (\ref{eq: potential KG}) does not depend on the function $g$ which appears in the black hole metric (\ref{eq: metric general}), in contrast, for the Dirac field the effective potentials (\ref{eq: potentials Dirac}) depend on the function $g$. We also note that in the massless limit the effective potentials (\ref{eq: potential KG}) and (\ref{eq: potentials Dirac}) go to zero. Thus in metric (\ref{eq: metric general}) the radial functions of the massless Klein-Gordon and Dirac fields satisfy free Schr\"odinger type equations. This fact is consistent with the absence of QNM for the massless Klein-Gordon and Dirac field in single horizon, asymptotically flat  two dimensional black holes. Furthermore when in \cite{Kettner:2004aw} the coupling function between the Klein-Gordon field and the dilaton is a constant, we get that in radial Schr\"odinger type equation the effective potential goes to zero. 

Finally, from the coupled equations (\ref{eq: Dirac coupled massive}) for the Dirac field we can get decoupled equations for the radial functions $R_1$ and $R_2$. The decoupled equations can be written in a similar form
\begin{equation} \label{e: radial Dirac general decoupled}
 g \frac{\dd^2 R_s}{\dd r^2} + \sqrt{g} \left( \frac{\dd \sqrt{g}}{\dd r} \right)\frac{\dd R_s}{\dd r} \pm i \omega \sqrt{g} \left( \frac{\dd }{\dd r} \frac{1}{\sqrt{f} } \right) R_s + \frac{\omega^2 R_s}{f} - m^2 R_s = 0, 
\end{equation} 
with $s=1,2$ and take the upper sign for $R_2$ and the lower sign for $R_1$. As far as we know the results of this Section have not been published.

\section{Quasinormal frequencies of the Dirac field in Witten black hole}
\label{s: Witten black hole}

A well studied two-dimensional spacetime is the uncharged Witten black hole \cite{Grumiller:2002nm}, \cite{Witten:1991yr}, \cite{Mandal:1991tz}, \cite{McGuigan:1991qp}, whose metric and dilaton are
\begin{align} \label{eq: metric Witten}
 \dd s^2  = \left(1-e^{-r}\right)\dd t^{2}-\frac{\dd r^{2}}{\left(1-e^{-r}\right)}, \qquad \qquad
\phi  = \phi_0 - \frac{r}{2},
\end{align} 
where $\phi_0$ is a constant.\footnote{Sometimes this black hole is called CGHS dilaton black hole \cite{Zelnikov:2008rg}.} This black hole is asymptotically flat and it has a horizon at $r_H = 0$ \cite{Witten:1991yr}, \cite{Mandal:1991tz}.

The propagation of test fields in the two-dimensional black hole (\ref{eq: metric Witten}) is analyzed in several papers, for example \cite{Zelnikov:2008rg}--\cite{Becar:2010zz}, \cite{Hsu:1992xt}--\cite{Frolov:1991ji}, mainly to study its classical and quantum stability. As the Witten black hole has a single horizon and is asymptotically flat, its QNM are the oscillations that satisfy the boundary conditions a) and b) of the previous section, (but see formula (\ref{eq: condition ratio omega}) below).

In Witten black hole for the coupled to scalar curvature massive Klein-Gordon field its QNF are calculated by Becar et al.\ in \cite{Becar:2007hu}, but as noted in \cite{LopezOrtega:2009zx}, there are additional QNF to those found in \cite{Becar:2007hu} (also these can be written in a simpler form \cite{LopezOrtega:2009zx}). The QNF of the non-minimally coupled to scalar curvature massive Klein-Gordon field are equal to \cite{Becar:2007hu}, \cite{LopezOrtega:2009zx}
\begin{eqnarray} \label{eq: QNF Klein-Gordon}
\omega &= -\frac{i}{4} \left[ 2n + 1 \pm  \sqrt{1-4\zeta} - \frac{4 m^2}{2n + 1 \pm \sqrt{1 - 4 \zeta}}\right] ,
\end{eqnarray} 
where $\zeta$ is the coupling constant to scalar curvature.  For the Witten black hole the QNF of the massless Dirac field are calculated in \cite{Becar:2010zz}, here we calculate the QNF of the Dirac field and comment on the existence of the QNF for the massless Dirac field  at the end of this Section. Furthermore notice that in \cite{Li:2001ct} the QNF of the charged Witten black hole \cite{McGuigan:1991qp} are calculated using the WKB method.

The metric of the Witten black hole takes the form (\ref{eq: metric general}) with 
\begin{equation} \label{eq: f g Witten}
 f(r) = g(r) = 1 - e^{-r} ,
\end{equation} 
therefore proposing separable solutions  as those of formula (\ref{eq: Dirac ansatz}), from (\ref{eq: Dirac coupled}) we find that in Witten black hole the Dirac equation  simplifies to
\begin{align} \label{eq: Dirac Witten radial}
(1 - e^{-r}) \frac{\dd R_{2} }{\dd r} + i\omega R_{2} &=  i m \sqrt{1 - e^{-r}} R_{1}, \nonumber\\
(1 - e^{-r}) \frac{\dd R_{1} }{\dd r} - i\omega R_{1} &= - i m \sqrt{1 - e^{-r}} R_{2}.
\end{align}

From this system of coupled equations for the radial functions $ R_{1}$ and $R_{2}$ we can get the decoupled equations (see also Eqs.\ (\ref{e: radial Dirac general decoupled}))
\begin{equation} \label{eq: radial equations Witten}
(1 - \textrm{e}^{-r}) \frac{\dd^{2} R_{s} }{\dd r^{2}} +   \frac{\textrm{e}^{-r}}{2} \frac{\dd R_{s} }{\dd r} - \frac{i \omega \nu  \textrm{e}^{-r}}{2 (1 - \textrm{e}^{-r}) }  R_{s} + \frac{\omega^{2}}{ 1 - \textrm{e}^{-r}} R_{s}  - m^{2} R_{s}=0,
\end{equation}
where
\begin{equation} \label{eq: nu}
 \nu = \left\{ \begin{array}{ll}
1 & \,\, \mbox{for  } s=2 ,\\
-1 & \,\, \mbox{for  } s=1 . \end{array} \right.
\end{equation} 
As for the Klein-Gordon field \cite{Becar:2007hu}, \cite{Frolov:2000jh} in Eqs.\ (\ref{eq: radial equations Witten}) we make the change of variable
\begin{equation} \label{eq: z r Witten}
 z = 1 - e^{-r},
\end{equation} 
to find that these equations transform into
\begin{equation} \label{eq: radial Witten z} 
\frac{\dd^{2} R_s }{\dd z^{2}} + \left(\frac{1}{2z} - \frac{1}{1-z}\right)\frac{\dd R_s }{\dd z} - \frac{i \omega \nu R_s}{2 z^{2} \left(1 -z \right) } + \frac{\omega^{2} R_s }{z^{2} \left(1 - z\right)^{2}} - \frac{m^{2} R_s }{z \left(1-z \right)^{2}}  =0.
\end{equation} 

To solve these differential equations we propose that the functions $R_s$ take the form  
\begin{equation} \label{eq: R ansatz}
R_s (z) = z^{\alpha_s} (1-z)^{\beta_s} F_s (z).
\end{equation}
If the parameters $\alpha_s$ and $\beta_s$ satisfy the equations
\begin{align} \label{eq: alpha beta Witten}
\alpha^{2}_s - \frac{\alpha_s}{2} + \omega^{2} - \frac{i\omega\nu}{2} = 0  , \qquad \qquad
 \beta_s^{2} +  \omega^{2} - m^{2} =0,
\end{align}
thus
\begin{align} \label{eq: alpha beta Dirac} 
\alpha_s =  \left\{ \begin{array}{l} i \omega \nu + \tfrac{1}{2},   \\ \\ - i \omega \nu, \end{array}\right. \qquad \quad
\beta_s  =  \left\{ \begin{array}{l}  \sqrt{ m^2 - \omega^2 } , \\ \\ -  \sqrt{ m^2 - \omega^2 } , \end{array}\right. 
\end{align}
then the functions $F_s$ are solutions of the differential equations 
\begin{align} \label{eq: radial F Witten}
z(1-z) \frac{\dd^{2}F_s}{\dd z^{2}} & +  \left(2 \alpha_s + \frac{1}{2} -  \left(2 \alpha_s + 2\beta_s + \frac{3}{2} \right) z \right)  \frac{\dd F_s}{\dd z} \nonumber \\
& + \left(2\omega^{2} - \alpha_s - \frac{i \omega \nu}{2}  - m^{2} - 2 \beta_s \alpha_s - \frac{\beta_s}{2}\right) F_s = 0.
\end{align}
We can transform the previous equations to the hypergeometric form \cite{Abramowitz-book}, \cite{Guo-book}
\begin{equation} \label{eq: hypergeometric equation}
 z(1-z)\frac{\dd^2 F}{\dd z^2} + (c -(a+b+1)z)\frac{\dd F}{\dd z} - a b F = 0 ,
\end{equation} 
with parameters
\begin{align} \label{eq: a b c Dirac}
a_s = \alpha_s + \beta_s +\tfrac{1}{2},\qquad \quad 
b_s =  \alpha_s + \beta_s, \qquad \quad 
c_s = 2 \alpha_s +\tfrac{1}{2}. 
\end{align}

In what follows we study in detail the component $\tilde{\Psi}_2$ ($\nu=1$) and choose the values of the parameters $\alpha_2$ and $\beta_2$ as
\begin{align} \label{eq: alpha beta Dirac 2}
\alpha_2 = i \omega + \tfrac{1}{2},  \qquad \qquad  \qquad
\beta_2 = \sqrt{m^{2}-\omega^{2}} .
\end{align}
If the parameter $c_2$ is not integer, then the solutions of Eq.\ (\ref{eq: radial F Witten}) with $s=2$ are \cite{Abramowitz-book}, \cite{Guo-book}
\begin{equation} \label{eq: }
F_2 (z) = \tilde{C}_{1} \, {}_{2} F_{1}(a_2,b_2;c_2;z) + \tilde{C}_{2} \, z ^{1-c_2}  {}_{2} F_{1}(a_2-c_2+1,b_2-c_2+1;2-c_2;z),
\end{equation}
with $\tilde{C}_1$, $\tilde{C}_2$ constants, ${}_{2} F_{1}(a,b;c;z)$ is the hypergeometric function and hence the radial function $R_2$ becomes
\begin{align} \label{eq: radial 2 Dirac}
R_2 &= \tilde{C}_{1} z^{ i\omega + 1/2} (1-z)^{\sqrt{m^{2} - \omega^{2}}} {}_{2}F_{1} (a_2,b_2;c_2;z) \nonumber \\ 
& + \tilde{C}_{2}z^{-i\omega}(1-z)^{\sqrt{m^{2} - \omega^{2}}} {}_{2}F_{1} (a_2-c_2+1,b_2-c_2+1;2-c_2;z).
\end{align}

Notice that the variable $z$ satisfies
\begin{equation}
 z = 0 \,\, \mbox{ at }  \,\, r = r_H, \qquad \qquad z \,\,  \to 1 \mbox{ as }  \,\, r \to \infty , 
\end{equation} 
therefore we find that near the horizon the radial function $R_2$ behaves as
\begin{equation} \label{eq: approximation horizon Dirac}
 R_2  \approx \tilde{C}_{1} z^{i\omega + 1/2 } + \tilde{C}_{2} z^{- i\omega} \approx \tilde{C}_{1} e^{ i \omega x + x/2} + \tilde{C}_{2} e^{-i\omega x},
\end{equation}
where $x$ is the tortoise coordinate (\ref{eq: tortoise coordinate general}) for the Witten black hole and is given by
\begin{equation} \label{eq: Witten tortoise}
 x = \ln(e^r - 1).
\end{equation} 

From (\ref{eq: Dirac ansatz}) we point out that the time dependence of the Dirac field is $\textrm{exp}(-i \omega t)$, therefore we obtain that in the near horizon expression (\ref{eq: approximation horizon Dirac}) for $R_2$, the first term is an outgoing wave, whereas the second term is an ingoing wave. Near the horizon the boundary condition of the QNM  imposes that the field must be purely ingoing, thus we take $\tilde{C}_1=0$. Hence the radial function $R_2$ takes the form 
 \begin{equation} \label{eq: radial 2 Dirac Witten}
R_2 = \tilde{C}_{2} z^{-i\omega }(1-z)^{\sqrt{m^{2}-\omega^{2}}} {}_{2}F_{1} (A_2, B_2; C_2; z),
\end{equation}
with
\begin{align} \label{eq: Dirac new parameters}
A_2 = a_2-c_2+1, \qquad \quad  
B_2 = b_2-c_2+1, \qquad \quad 
C_2 &= 2-c_2 .
\end{align}

As is well known, when $c$ is not a negative integer and $c - a -b$ is not an integer the hypergeometric function ${}_{2}F_{1}(a,b;c;z)$ satisfies Kummer's property \cite{Abramowitz-book}, \cite{Guo-book}
\begin{align} \label{eq: hypergeometric property z 1-z}
{}_2F_1(a,b;c;z) &= \frac{\Gamma(c) \Gamma(c-a-b)}{\Gamma(c-a) \Gamma(c - b)} {}_2 F_1 (a,b;a+b+1-c;1-z) \nonumber \\
&+ \frac{\Gamma(c) \Gamma( a + b - c)}{\Gamma(a) \Gamma(b)} (1-z)^{c-a -b} {}_2F_1(c-a, c-b; c + 1 -a-b; 1 -z),
\end{align} 
where $\Gamma(a)$ denotes the gamma function. 

Taking into account Kummer's property we write the radial function $R_2$ as
\begin{align} \label{eq: radial Dirac Kummer}
R_2 & = \tilde{C}_{2} z^{- i\omega} (1-z)^{\sqrt{m^{2} - \omega^{2}}} \frac{\Gamma(C_2) \Gamma(C_2-A_2-B_2)}{\Gamma(C_2-A_2) \Gamma(C_2-B_2)} {}_{2}F_{1} (A_2,B_2;A_2+B_2-C_2+1;1-z)  \nonumber \\
& + \tilde{C}_{2} z^{- i\omega} (1-z)^{-\sqrt{m^{2} - \omega^{2}}} \frac{\Gamma(C_2) \Gamma(A_2+B_2-C_2)}{\Gamma(A_2) \Gamma(B_2)} \nonumber \\
& \qquad \qquad \qquad \qquad \qquad \times {}_{2}F_{1} (C_2-A_2,C_2- B_2;C_2-A_2-B_2+1;1-z).
\end{align}
As $r \to + \infty$ ($z \to 1$) the previous expression simplifies to (note that as $z \to 1$, we can use the approximation $1-z \approx e^{-x}$)
\begin{align} \label{eq: approximation radial infinity}
R_2 & \approx  \frac{\Gamma(C_2)\Gamma(C_2-A_2-B_2)}{\Gamma(C_2-A_2)\Gamma(C_2-B_2)} e^{-\sqrt{m^{2}-\omega^{2}} x} + \frac{\Gamma(C_2)\Gamma(A_2+B_2-C_2)}{\Gamma(A_2)\Gamma(B_2)}e^{\sqrt{m^{2}-\omega^{2}} x} .
\end{align} 

To determine precisely what is the ingoing or outgoing wave, we assume that
\begin{equation} \label{eq: condition ratio omega}
 \frac{\sqrt{\omega^{2} - m^{2}} }{\omega} > 0,
\end{equation} 
to find that as $r \to + \infty$ the first term of (\ref{eq: approximation radial infinity}) is an ingoing wave, whereas the second is an outgoing wave \cite{Ohashi:2004wr}. The boundary condition of the QNM imposes that we have only outgoing waves at infinity. Taking into account the properties of the gamma function we must satisfy the condition
\begin{equation} \label{eq: Dirac conditions QNF}
C_2-A_2= -n, \qquad \textrm{or} \qquad C_2-B_2 = -n, \qquad n = 0, 1, 2, \dots 
\end{equation}

From the previous equations we find that for the component $\tilde{\Psi_2}$ of the Dirac field its QNF are equal to
\begin{align} \label{eq: QNF Dirac 2}
\omega_{2} = - \frac{i}{2n} \left(n^{2} - m^{2} \right), \qquad \qquad
\omega_{2} = - \frac{i}{2\left(n+\frac{1}{2}\right)} \left(\left(n + \tfrac{1}{2} \right)^{2} -m^{2} \right).
\end{align}
Notice that for the first set of the previous QNF we must take $n=1,2,3,\dots$. Using a similar method we find that for the component $\tilde{\Psi_1}$ of the Dirac field its QNF are
\begin{align} \label{eq: QNF Dirac 1}
\omega_{1} = -\frac{i}{2(n+1)} \left( (n+1)^{2} - m^{2} \right),  \qquad 
\omega_{1} = -\frac{i}{2\left(n +\tfrac{1}{2}\right)} \left(\left( n + \tfrac{1}{2} \right)^{2} -m^{2}\right) ,
\end{align}
with $n=0,1,2,\dots$.

Thus for the components $\tilde{\Psi}_1$ and $\tilde{\Psi}_2$ we find that their QNF are purely imaginary as those of Refs.\ \cite{Fernando:2003ai}--\cite{Zelnikov:2008rg}. We also note that $\tilde{\Psi}_1$ and $\tilde{\Psi}_2$ have one set of QNF that coincides with one set of the QNF for the minimally coupled Klein-Gordon field of the same mass (\ref{eq: QNF Klein-Gordon}), but both components $\tilde{\Psi}_1$ and $\tilde{\Psi}_2$ have an additional set of QNF, and this additional set is different from the QNF of the minimally coupled massive Klein-Gordon field. This result illustrates that even in two dimensions the Klein-Gordon and Dirac fields have different responses to the gravitational fields. 

We say that the QNM are stable when their amplitudes decrease as time increases. From time dependence taken in formulas (\ref{eq: separation KG}) and (\ref{eq: Dirac ansatz}) we obtain that the QNM are stable for $\omega_I < 0$, where $\omega_I$ denotes the imaginary part of the QNF. For the Witten black hole we find that the QNM of the component $\tilde{\Psi}_1$ are stable when
\begin{equation} \label{eq: stability Dirac 1}
 n > m-\tfrac{1}{2}, 
\end{equation} 
and the QNM of the component $\tilde{\Psi}_2$ are stable when
\begin{equation} \label{eq: stability Dirac 2}
 n > m \qquad \textrm{for} \qquad  n \geq 1 \qquad \textrm{and} \qquad \tfrac{1}{2} > m \qquad \textrm{for} \qquad n = 0.
\end{equation} 
For the Klein-Gordon field in Witten black hole the stability of its QNM  is studied in \cite{Becar:2007hu}.

Thus for the Dirac field we may expect that at least the fundamental QNM is unstable, and depending on the value of the mass additional QNM may be unstable. Nevertheless we recall that in the method used to calculate QNF (\ref{eq: QNF Dirac 2}) and (\ref{eq: QNF Dirac 1}) we impose the condition (\ref{eq: condition ratio omega}), thus we must verify whether these QNF satisfy this condition. For the component $\Psi_2$, from the first set of QNF (\ref{eq: QNF Dirac 2}) follows
\begin{equation}
 \frac{\sqrt{\omega^{2} - m^{2}} }{\omega} = \frac{n^2 + m^2}{n^2 - m^2} > 0, \qquad \textrm{for}  \qquad n > m,  \qquad (n \geq 1),
\end{equation} 
and for the second set of QNF (\ref{eq: QNF Dirac 2}) we get
\begin{equation}
 \frac{\sqrt{ \omega^{2} - m^{2}} }{\omega} = \frac{(n+1/2)^2 + m^2}{(n+1/2)^2 - m^2} > 0, \qquad \textrm{for}  \qquad n + \tfrac{1}{2} > m, \qquad (n \geq 0).
\end{equation} 
Hence only the stable QNM (those that fulfill the stability condition (\ref{eq: stability Dirac 2})) satisfy our assumption (\ref{eq: condition ratio omega}). 

So this fact points out that the modes that fail to keep the stability condition (\ref{eq: stability Dirac 2}), in addition they do not satisfy condition (\ref{eq: condition ratio omega}) that we use to calculate the QNM of the Witten black hole. A similar result is valid for QNF (\ref{eq: QNF Dirac 1}) of the component $\tilde{\Psi}_1$. We think that these modes must be studied in more detail, because in \cite{Becar:2007hu} for the Klein-Gordon field  are found some unstable QNM. Moreover the results by Azreg-Ainou \cite{AzregAinou:1999kb} point out that the Witten black hole has some classical instabilities. Nevertheless for the Dirac field we obtain that the possible unstable modes do not fulfill our assumption (\ref{eq: condition ratio omega}).

Since the Witten black hole is a single horizon, asymptotically flat two-dimensional black hole and its tortoise coordinate (\ref{eq: Witten tortoise}) satisfies conditions (\ref{eq: conditions tortoise}), from the results explained in Section \ref{s: QNM massless} we expect that the minimally coupled massless Klein-Gordon and Dirac fields do not have well defined QNF. Nevertheless analytical expressions for the QNF of the massless Dirac field are presented in Ref.\ \cite{Becar:2010zz} (see formula (37) in \cite{Becar:2010zz}). Thus on the existence of the QNF for the massless Dirac field in  Witten black hole our results of Section \ref{s: QNM massless} contradict those of \cite{Becar:2010zz}.

To solve this issue we note that in Ref.\ \cite{Becar:2010zz} there is an inconsistency. To write solutions (29) of \cite{Becar:2010zz} Becar et al.\ assume that their parameter $c$ is not an integer \cite{Guo-book}. Furthermore to write their equation (34) they assume that their parameter $c$ is not a negative integer and that their quantity $c-a-b$ is not an integer because they use Kummer's property of the hypergeometric function \cite{Guo-book}. From the expressions for the QNF of the massless Dirac field given in \cite{Becar:2010zz} (see their formula  (37)), and using the expressions for the parameters $a$, $b$, and $c$ that appear in \cite{Becar:2010zz} (see their formulas (26)-(28)) we find that for frequencies (37) of \cite{Becar:2010zz}, their quantities $c$ and $c-a-b$ are integers. This fact shows that Becar et al.\ contradict their assumptions and therefore the frequencies that they give in \cite{Becar:2010zz} are not QNF of the massless Dirac field.

The previous comments imply that the Witten black hole does not have well defined QNM when the quantity $c$ of Ref.\ \cite{Becar:2010zz} is not an integer. It is convenient to extend the analysis of \cite{Becar:2010zz} to integral values of $c$. For $c$ equal to an integer it is expected that one solution of Eq.\ (25) in \cite{Becar:2010zz} involves logarithmic terms \cite{Guo-book}. Nevertheless for this problem the solutions can be found in a simple closed form. 

First we note that to obtain decaying modes we must take $c=N$ with $n \in \mathbb{Z}$, $N \leq 0$. Furthermore when $c=N$ we note that the quantities $a$, $b$, and $c-a-b$ of Eq.\ (25) in \cite{Becar:2010zz} are integers and we notice that for this value  of $c$ one of the quantities $a$, $b$ is always a negative integer. As a consequence we expect the simplification of the solutions \cite{Guo-book}.

In fact, when $c$ is a non-positive integer we find that Eq.\ (25) of \cite{Becar:2010zz} transforms into\footnote{In formulas (\ref{e: Becar hypergeometric})--(\ref{e: Becar solutions phi two}) we follow the conventions of Ref.\ \cite{Becar:2010zz}.}
\begin{equation} \label{e: Becar hypergeometric}
 z(1-z)g^{\prime \prime}(z) + (c-2cz)g^{\prime}(z) - c(c-1)g(z)=0,
\end{equation} 
whose solutions are
\begin{equation} \label{e: Becar solutions g}
 g(z)=\tilde{C}_1 z^{1-c} + \tilde{C}_2 (1-z)^{1-c},
\end{equation} 
where $\tilde{C}_1$ and $\tilde{C}_2$ are constants (as previously).

Therefore the component $\phi_2(z)$ of formula (24) in \cite{Becar:2010zz} is equal to
\begin{equation} \label{e: Becar solutions phi two}
 \phi_2(z)=z^{1/4}(\tilde{C}_1 z^{i \omega} (1-z)^{-i \omega} + \tilde{C}_2 z^{- i \omega} (1-z)^{i \omega} ).
\end{equation} 
From this expression for $\phi_2(z)$ we get that the first term is an outgoing wave near the horizon and at infinity. The second term is an ingoing wave near the horizon and at infinity. Hence these solutions do not satisfy the boundary conditions of the QNM, and also when the quantity $c$ is an integer we do not find QNF for the massless Dirac field in Witten black hole.

Thus according to the results of Section \ref{s: QNM massless} we believe that in Witten black hole the minimally coupled massless Klein-Gordon and Dirac fields do not have well defined QNF. Notice that for the Dirac field propagating in Witten black hole for QNF (\ref{eq: QNF Dirac 2}) and (\ref{eq: QNF Dirac 1}) the limits $m \to 0$ are well defined, nevertheless the frequencies we get in these limits are not QNF.

For the coupled to scalar curvature massive Klein Gordon field propagating in Witten black hole the effective potential is \cite{Frolov:2000jh}
\begin{equation} \label{eq: potential KG Witten}
V = \frac{m^2 (\textrm{e}^r - 1)}{\textrm{e}^r} + \frac{\zeta (\textrm{e}^r - 1) }{\textrm{e}^{2r}} =\frac{m^2 e^x}{e^x+1} + \frac{\zeta e^x}{(e^x +1)^2}.
\end{equation} 
Following Chandrasekhar \cite{Chandrasekhar book} we find that in Witten black hole for the Dirac field its effective potentials take the form 
\begin{equation} \label{eq: potential Dirac Witten}
V_\pm = m^2 (1-e^{-r}) \mp \frac{m}{2}\frac{(1-e^{-r})^{1/2}}{e^r} = \frac{m^2 e^x}{e^x + 1} \mp \frac{m}{2} \frac{e^{x/2}}{(e^x + 1)^{3/2}}.
\end{equation} 
We point out that for the minimally coupled fields  the effective potentials (\ref{eq: potential KG Witten}) and (\ref{eq: potential Dirac Witten}) go to zero as the mass of the field goes to zero (it is expected from the results of Section \ref{s: QNM massless}). Furthermore notice that the effective potentials $V$ and $V_\pm$ satisfy
\begin{equation}
\lim_{x \to -\infty} V = \lim_{x \to -\infty} V_\pm = 0, \qquad \qquad \lim_{x \to + \infty} V = \lim_{x \to + \infty} V_\pm = m^2. 
\end{equation}

\section{Summary}
\label{s: Summary}

We find that the minimally coupled Klein-Gordon and Dirac fields do not have modes that satisfy the standard boundary conditions for the QNM of single horizon, asymptotically flat two-dimensional black holes. Thus if we are interested in calculating the QNF of the Klein-Gordon and Dirac fields in a black hole of this family, then we must analyze the massive fields (as in the previous section and \cite{Zelnikov:2008rg}, \cite{Becar:2007hu}, \cite{LopezOrtega:2009zx}), fields non-minimally coupled to scalar curvature (as in \cite{Zelnikov:2008rg}, \cite{Becar:2007hu}) or to dilaton (as in \cite{Kettner:2004aw}). This result shows that for two-dimensional black holes when we know the QNF of a massive field and take the massless limit,  we must verify whether the frequencies obtained in this limit are QNF. We believe that the consequences of this result deserve further research.

In \cite{Becar:2010zz} for the Witten black hole  Becar et al.\ assert that the QNM of the massless Dirac field are stable, but based on the results of Section \ref{s: QNM massless} we believe that in the massless case the QNF of this fermion field are not well defined, and we also show that there are inconsistencies in the method used in Ref.\ \cite{Becar:2010zz}. Therefore the conclusion of Ref.\ \cite{Becar:2010zz} on the stability of the QNM for the massless Dirac field in Witten black hole is not valid, because the QNM are not defined in the massless limit. Furthermore  for both components of the Dirac field we find that the modes not satisfying the stability conditions (\ref{eq: stability Dirac 1}) and (\ref{eq: stability Dirac 2}), at the same time they do not fulfill the assumption (\ref{eq: condition ratio omega}) and it is probable that these modes are not QNM. Notice that for the Klein-Gordon field propagating in Witten black hole some classical and quantum instabilities are known  (see Refs.\ \cite{Zelnikov:2008rg}, \cite{Becar:2007hu}, \cite{LopezOrtega:2009zx}, \cite{AzregAinou:1999kb}, \cite{Frolov:2000jh}, \cite{Frolov:1991ji}).

\section{Acknowledgments}

We thank to Dr.\ R.\ Cordero Elizalde his interest and support of this project. This work was supported by CONACYT M\'exico, SNI M\'exico, EDI-IPN, COFAA-IPN, and Research Projects SIP-20110729 and SIP-20111070. I.\ Vega-Acevedo acknowledges financial support from CONACYT research grant no.\ J1-60621-I.

\end{document}